\documentclass{aa}
\usepackage{graphicx}
\usepackage{natbib}
\usepackage{color}
\usepackage{float}
\usepackage{txfonts}
\usepackage{pdflscape}

\begin{document}

\title{Light-time effect detected in fourteen eclipsing binaries}

\author{Zasche, P.~\inst{1},
 Uhla\v{r}, R.~\inst{2},
 Svoboda, P.~\inst{3},
 Caga\v{s}, P.~\inst{4},
 Ma\v{s}ek, M.~\inst{5}
 }

 \institute{
  $^{1}$ Astronomical Institute, Charles University, Faculty of Mathematics and Physics, CZ-180~00, Praha 8, V~Hole\v{s}ovi\v{c}k\'ach 2, Czech Republic,  \email{zasche@sirrah.troja.mff.cuni.cz}\\
  $^{2}$ Private Observatory, Poho\v{r}\'{\i} 71, CZ-254 01 J\'{\i}lov\'e u Prahy, Czech Republic, \\
  $^{3}$ Private Observatory, V\'ypustky 5, CZ-614 00, Brno, Czech Republic,\\
  $^{4}$ BSObservatory, Modr\'a 587, CZ-760~01, Zl\'{\i}n, Czech Republic,\\
  $^{5}$ Institute of Physics, Czech Academy of Sciences, Na Slovance 1999/2, CZ-182~21, Praha 8, Czech Republic.\\
  }

\titlerunning{Study of 14 eclipsing binaries}
\authorrunning{Zasche et al.}

  \date{Received \today; accepted ???}

\abstract{The available minima timings of 14 selected eclipsing binaries (V1297~Cas, HD~24105, KU~Aur,
GU~CMa, GH~Mon, AZ~Vel, DI~Lyn, DK~Her, GQ~Dra, V624~Her, V1134~Her, KIC~6187893, V1928~Aql, V2486~Cyg)
were collected and analyzed. Using the automatic telescopes, surveys, and satellite data, we derived
more than 2500 times of eclipses, accompanied with our own ground-based observations. These data were
used to detect the period variations in these multiple systems. The eclipse timing variations (ETVs)
were described using the third-body hypothesis and the light-time effect. Their respective periods were
derived as 2.5, 16.2, 27, 20, 64, 5.6, 22, 115, 27, 42, 6.9, 11.2, 4.1, and 8.4~ years for these
systems, respectively. The predicted minimal mass of the third body was calculated for each of the
systems, and we discuss here their prospective detectability. The light curves of HD~24105, GH~Mon,
DK~Her, V1134~Her, KIC~6187893, V1928~Aql, and V2486~Cyg were analyzed using the PHOEBE program,
resulting in physical parameters of the components. Significant fractions of the third light were
detected during the light-curve analysis, supporting our hypothesis of the triple-star nature of all
these systems. The majority of these systems (nine out of 14) were already known as visual doubles. Our
study shifts them to possible quadruples, what makes them even more interesting.}

\keywords {stars: binaries: eclipsing -- stars: fundamental parameters}

\maketitle

\section{Introduction} \label{intro}

The role of classical eclipsing binaries (EBs) in modern astrophysics is undisputable. Surprisingly,
after a century of their intensive research, they still represent the most general method used to
derive the basic stellar properties with the highest accuracy (see e.g.,
\citealt{2012ocpd.conf...51S}). Quantities such as masses, radii, or luminosities can be derived
with unprecedented precision at the level of about 1\% only, also yielding the distance at about
the same precision: nowadays even outside of our own Galaxy \citep{2019Natur.567..200P}.

Besides this role of EBs as independent distance indicators, their
physical parameters should obviously be used for calibrating the existing stellar evolution models
\citep{2010A&ARv..18...67T}. Moreover, the EBs with their precise eclipses can also serve a
different purpose: as ideal tics measuring precise time intervals. Hence, studying the
deviations from the predicted ephemerides via studying the delays with eclipse timing
variations (ETVs) is nowadays a classical method, which can be used for revealing some of the hidden
properties of the binaries themselves \citep{2005ASPC..335....3S}.

This method of ETV analysis usually employs a standard light-time effect hypothesis of additional
body orbiting around a barycenter (hereafter LITE, see e.g., \citealt{1959AJ.....64..149I}, and
\citealt{Mayer1990}). Such an approach is substantiated by the fact that if the third bodies move on
sufficiently long orbits, we can assume that the whole system is almost like a two-body problem.
However, when the orbits are closer to each other (and the ratio of periods ${p_3\over P}
\lesssim 100$), then the dynamical interaction of the third body on the inner double cannot be
neglected. These tight triples show many different ETV signals, depending on their masses,
orientations of the orbits, eccentricities, and so on. See, for example, \cite{2003A&A...398.1091B} or
\cite{2011A&A...528A..53B}. However, such tight multiples can usually only be detected with
dense data obtained with high cadency over a certain time interval, like in
\cite{2015MNRAS.448..946B}. Moreover, these close third bodies are also able to cause other
effects like orbital precession or even inclination changes of the inner eclipsing double.

On the other hand, many studies on potential triples discovered via ETV applying the LITE
hypothesis were later found to be unrealistic, when new data were available. Therefore, we have to be cautious of
our interpretation, and our result is still rather a hypothesis waiting to be proven by
other methods, such as spectroscopy. Besides the LITE, there are also some alternative mechanisms
present that cause the ETVs, especially in contact systems, like the modulation through spots and
magnetic activity cycles: for example, the so-called Applegate mechanism \citep{1992ApJ...385..621A}.

The importance of the third bodies on the origin and subsequent evolution of the binaries
(also including the role of the Kozai cycles) has been discussed elsewhere: for example,
\cite{2001ApJ...562.1012E} or \cite{2008MNRAS.389..925T}. However, the origin of these
hierarchical multiples is still being discussed and remains an open question
\citep{2019ApJ...883...22C}.

\section{Methods}  \label{methods}

Our method of detection and characterization of the sample binaries uses a classical approach
assuming only one body on a quite distant orbit, meaning there is no dynamical interaction, and only the LITE term was
used for ETV analysis. With this method, we would be able to find preferably the bodies with periods of
a few years to decades, and with LITE semiamplitudes of the order of 0.01~day (quite similarly to our
former analysis, \citealt{2014AJ....147..130Z}). This limitation mainly comes from the fact that the
data cadence is typically limited, and the precision of the individual times of eclipses critically
depends on the quality of the light curve (LC) and depth of a particular eclipse.

The times of eclipses for the ETV analysis were collected from published papers. Additionally to the published eclipses, we also computed the new ones from the available data. These
were our new dedicated observations of these targets (from ground-based observatories, using
typically small amateur telescopes), as well as the rich databases of photometric observations of
these binaries from existing archives. The PDR code \citep{2019CoSka..49..132Z} was used to
obtain some of these data. These datasets were the following:

\begin{itemize}
 \item HIP: the Hipparcos satellite, observing in special $H_p$ filter, between 1989 and 1993 \citep{1997A&A...323L..49P}.
 \item NSVS: the ROTSE-I experiment, unfiltered photometry, with a time span from 1999 to 2000 \citep{2004AJ....127.2436W}.
 \item OMC: five-cm camera onboard the INTEGRAL satellite, observing in $V$ filter since 2002 \citep{2003A&A...411L.261M}.
 \item ASAS: All Sky Automated Survey, observing since 1997 in $V$ and $I$ filters \citep{2002AcA....52..397P}.
 \item ASAS-SN: ASAS for supernovae, 24 telescopes, $V$ and $g$ filters \citep{2014ApJ...788...48S,2017PASP..129j4502K}.
 \item SuperWASP: 20-cm telescopes, using special filters, observing since 2004 \citep{2006PASP..118.1407P}.
 \item Pi of the sky: small robotic cameras, observing since 2004, unfiltered \citep{2005NewA...10..409B}.
 \item CRTS: Catalina Survey, 70-cm telescope, observing since 2007, unfiltered \citep{2009ApJ...696..870D}.
 \item Kepler: Kepler satellite, 95-cm telescope, observed from 2009 to 2018 in special filter \citep{2010Sci...327..977B}.
 \item TESS: TESS satellite, 10-cm diameter, observing since 2018 in special filter \citep{2015JATIS...1a4003R}.
 \item KWS: Kamogata/Kiso/Kyoto wide-field survey, observing in $BVI_c$ filters \citep{KWS}.
 \item MASCARA: small 17-mm wide-field cameras,  observing since 2017, unfiltered \citep{2018A&A...617A..32B}.
 \item ZTF: the Zwicky Transient Facility, 120-cm telescope, $g$, and $r$ filters \citep{2019PASP..131a8003M}.
\end{itemize}

For the derivation of times of mid-eclipse, our automatic fitting procedure (AFP,
\citealt{2014A&A...572A..71Z}) was used. It uses an LC template and phased LCs at particular time
intervals when the phase coverage of both minima is sufficient. Therefore, the cadency of these derived
minima highly depends on the particular survey, its cadency, and the precision of individual
photometric data points.

For the systems where no previous LC solution exists, we also performed an LC analysis. This was typically based on the best available LC for that particular system, meaning
from the source (from the aforementioned list) that provides the best phase coverage as well as
the lowest scatter of the individual observations. The individual cases are discussed in more
detail below.

For the LC solution, we used {\sc PHOEBE}, ver 0.32svn \citep{2005ApJ...628..426P} software. It
uses a well-known \cite{1971ApJ...166..605W} algorithm, with its later modifications. We usually
assumed fixed coefficients of albedo and gravity brightening, according to the assumed temperature.
The mass ratio was kept fixed during the whole analysis due to missing radial velocities. Hence,
only the primary temperature was a necessary input for {\sc PHOEBE} in the beginning. This was
taken as a typical effective temperature for a particular spectral type (as given in Table
\ref{InfoSystems}), or was assumed according to its photometric index following the tables by
\cite{2013ApJS..208....9P}, with recent updates available
online.\footnote{\tiny{www.pas.rochester.edu/$\sim$emamajek/EEM$\_$dwarf$\_$UBVIJHK$\_$colors$\_$Teff.txt}}
The main-sequence assumption was naturally used as a simplest approach due to our having only very
limited information about these binaries.

\section{The individual systems under analysis}

The selected eclipsing binaries in our sample were required to have the following: a range of
magnitudes between six and 13 and an orbital period of up to four days; a times-of-minima dataset
sufficiently large for a period analysis; an obvious variation in the $O-C$ diagram; and a third-body
hypothesis never published before.

Using these criteria, 14 systems were found to be suitable for the present analysis. We focus
on individual systems one-by-one in the following subsections. Basic information about these stars
is summarized in Table \ref{InfoSystems}, which presents the position on the sky as well as
photometric indices and published spectral types, if available.

\begin{table*}[h!]
\caption{Relevant information for the analyzed systems.}  \label{InfoSystems}
 \tiny
\begin{tabular}{lccccccccccc}
   \hline\hline\noalign{\smallskip}
  System      & Other ID          &     RA    & DE &$V_{\rm max}[\mathrm{mag}]$&$(J-H)[\mathrm{mag}]^{B}$ & $(B-V)[\mathrm{mag}]^{C}$& Published Sp.Type \\
  \hline\noalign{\smallskip}
  \object{V1297 Cas} & HD 232102        & 00 06 37.38 & +55 27 21.72 & 9.41 &  0.398          &  0.984          & K0 \citep{1927AnHar.100...33C} \\
   \object{HD 24105} & EPIC 211160717   & 03 51 21.23 & +25 38 08.39 & 9.43 &  0.363          &  0.697          & G5 \citep{1975ascp.book.....H} \\
    \object{KU Aur}  & TYC 2422-20-1    & 06 28 04.38 & +30 23 34.01 &11.77 &  0.347          &  0.655          & F8 \citep{2015RAA....15.1095L} \\
    \object{GU CMa}  & HD 52721         & 07 01 49.51 & -11 18 03.32 & 6.59 &  0.087          &  0.038          & B2 \citep{1968PASP...80..197G} \\
    \object{GH Mon}  & GSC 04818-02919  & 07 04 41.44 & -02 28 19.01 &12.70 &  0.149          &                 & F6 \citep{2015RAA....15.1095L} \\
    \object{AZ Vel}  &                  & 08 22 51.89 & -44 25 45.39 &12.72 &  0.272          &                 &                                \\
    \object{DI Lyn}  & A Hya            & 09 35 22.51 & +39 57 47.78 & 6.76 &  0.120          &  0.362          & F2 \citep{1945PDDO....1..311Y} \\
    \object{DK Her}  & HD 155700        & 17 12 43.95 & +13 11 03.77 &10.87 &  0.136          &  0.306          & A1 \citep{2015RAA....15.1095L} \\
    \object{GQ Dra}  & HD 158260        & 17 25 29.44 & +51 29 35.10 & 9.14 &  0.112          &  0.198          & A3 \citep{1988AaAS...74..449R} \\
  \object{V624 Her}  & HD 161321        & 17 44 17.25 & +14 24 36.24 & 6.22 &  0.034          &  0.226          & A3 \citep{1984AJ.....89.1057P} \\
  \object{V1134 Her} & GSC 01031-01766  & 18 28 14.49 & +12 19 51.06 &12.70 &  0.207          &                 &                                \\
 \object{KIC 6187893}& TYC 3128-1653-1  & 19 02 04.71 & +41 33 00.34 &11.84 &  0.451          &  1.125          & G2 \citep{2015RAA....15.1095L} \\
  \object{V1928 Aql} & HD 180848        & 19 17 44.84 & +08 46 53.79 & 9.75 &  0.220          &  0.521          & A3 \citep{1988AaAS...74..449R} \\
  \object{V2486 Cyg} & GSC 03173-01826  & 21 16 59.44 & +40 19 56.71 & 9.50 &  0.101          &  0.229          & A2 \citep{1988AaAS...74..449R} \\
 \noalign{\smallskip}\hline
\end{tabular}
 \\
 \scriptsize Note: [B] - 2MASS catalog by \cite{2006AJ....131.1163S}; [C] - based on the Tycho catalog by \cite{2010PASP..122.1437P}.
\end{table*}

\subsection{V1297 Cas}

The first system in our sample is V1297~Cas (= HD~232102 = NSVS~1557555), which is also the one
with the latest spectral type classification (K0 according to \citealt{1927AnHar.100...33C}). It
was quite recently studied photometrically, when \cite{2017NewA...57...37B} published their $BVI_c$
light curves. Their detailed analysis revealed that this W~UMa-type contact binary with an orbital
period of about 0.27 days has an inclination of about 85$^\circ$ and mass ratio $q=0.55$.
What was quite surprising is the fact that they also detected quite a large fraction of the third
light in their LC solution. There was about 58\% in $B$, while about 49\% in $I_c$, together with a hot
spot and asymmetric LC.

\begin{figure}
  \centering
  \includegraphics[width=0.48\textwidth]{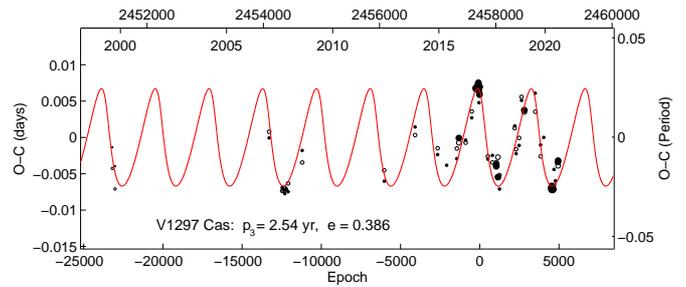}
  \caption{Period analysis of V1297~Cas. The open circles stand for the secondary, while the filled dots
  for the primary minima. The larger the symbol, the higher the precision. The solid line represents the
  predicted third-body variation using the light-time effect hypothesis.}
  \label{FigV1297Cas_LITE}
\end{figure}

According to these findings, the indication of the third component in the system would not be so
surprising. Therefore, we collected all available times of minima: from \cite{2017NewA...57...37B},
as well as from other publications. Moreover, we also derived more than 170 additional eclipse timings
based on other photometric sources. These were: ASAS-SN, KWS, TESS, NSVS, and SuperWASP. These new
minima times greatly accompanied the existing data, clearly revealing the periodic
variation. The ETV signal together with the fit is plotted in Fig. \ref{FigV1297Cas_LITE}. The LITE
hypothesis was used, resulting in the parameters given in Table \ref{LITEparam}. The variation with
about 2.5-yr periodicity is clearly visible here.

From the LITE parameters, we also tried to estimate the properties of the distant body. Mass
function of the third body resulted in about 0.3~M$_\odot$. This indicates that the third body
should have at least $M_{3,min} = 1.32$~M$_\odot$ (assuming the coplanar orbits). On the other
hand, \cite{2017NewA...57...37B} estimated the spectral type of the third body as G7. To conclude,
we can certainly state that the third component is the dominant body in the system. From the GAIA
\citep{2018A&A...616A...1G} parallax, we can also compute the predicted angular separation of the
third component on the sky for a prospective interferometric detection. Here, it resulted in about
36~mas, which is within the capabilities of modern instruments, hence it is realistic to assume that it will be
detected soon.

\begin{table*}
 \centering
 \scriptsize
  \caption{Final parameters of the LITE orbits, corresponding to the LITE fits plotted in the figures. Seven parameters from the first part of the table were
  fit, while the last three columns were only computed from these parameters.}  \label{LITEparam}
  \begin{tabular}{@{}c | c c c c c c c | c c c@{}}
\hline
 Parameter  &    $JD_0 $     &      $P$      &   $p_3$    &      $A$   &    $T_0$     &  $\omega$  &    $e$     & $f(m_3)$    & $M_{3,min}$ &  $a$       \\
 Unit       &  HJD-2450000   &      [d]      &    [yr]    &     [day]  &  HJD-2450000 &   [deg]    &            & [M$_\odot$] & [M$_\odot$] & [mas]      \\ \hline
 V1297 Cas  & 7722.4411 (6)  &  0.272511 (2) & 2.54 (0.02)& 0.0068 (3) & 8701.5 (24.0)&144.0 (10.1)& 0.386 (56) & 0.299 (3)   & 1.32 (2)    & 35.7 (1.6) \\
 HD 24105   & 4214.7294 (32) &  1.262921 (11)& 16.2 (3.2) & 0.0063 (22)& 5982.0 (1426)& 70.3 (28.8)& 0.247 (49) & 0.005 (1)   & 0.29 (1)    & 40.6 (10.2)\\
 KU Aur     & 6700.3096 (14) &  1.319579 (3) & 26.6 (0.7) & 0.0099 (11)& 6241.3 (914) &  0.1 (1.5) & 0.352 (110)& 0.009 (1)   & 0.28 (1)    & 11.3 (4.8) \\
 GU CMa     & 5262.4617 (21) &  1.610132 (14)& 19.7 (0.9) & 0.0318 (14)& 1308.6 (374) & 88.9 (12.0)& 0.880 (78) & 0.433 (15)  & 4.57 (9)    & 40.3 (16.9)\\
 GH Mon     & 3407.2712 (29) &  0.707184 (1) & 64.2 (6.8) & 0.0076 (20)& 3399.3 (1278)&245.3 (43.2)& 0.799 (44) & 0.001 (1)   & 0.17 (3)    & 17.8 (4.0) \\
 AZ Vel     & 2805.4518 (47) &  0.775776 (46)& 5.65 (0.71)& 0.0054 (19)& 1957.2 (326) &167.8 (86.3)& 0.261 (17) & 0.029 (8)   & 0.58 (9)    &  5.6 (2.1) \\
 DI Lyn     & 4591.2565 (32) &  1.681566 (3) & 22.3 (3.5) & 0.0113 (30)& 5574.5 (392) &302.4 (32.7)& 0.797 (159)& 0.020 (4)   & 0.57 (4)    &146.5 (23.5)\\
 DK Her     & 5730.5182 (132)&  1.942291 (2) &115.2 (14.9)& 0.0710 (85)& 5198.6 (6200)&358.2 (17.1)& 0.087 (18) & 0.142 (13)  & 1.70 (13)   & 73.9 (9.2) \\
 GQ Dra     & 6052.3586 (44) &  0.765903 (1) & 27.4 (0.7) & 0.0065 (29)& 9445.1 (338) &180.2 (18.0)& 0.746 (47) & 0.007 (2)   & 0.42 (18)   & 26.4 (11.7)\\
 V624 Her   & 7203.8045 (17) &  3.894983 (2) & 42.2 (4.0) & 0.0069 (17)& 5025.2 (1560)&  0.6 (21.2)& 0.695 (117)& 0.003 (1)   & 0.38 (14)   &142.5 (40.1)\\
 V1134 Her  & 6497.4564 (9)  &  0.602934 (1) & 6.86 (0.09)& 0.0090 (9) & 3522.0 (82)  &250.1 (14.4)& 0.749 (125)& 0.090 (6)   & 1.11 (8)    &  5.7 (0.6) \\
 KIC 6187893& 4954.0702 (3)  &  0.789183 (2) & 11.2 (0.2) & 0.0152 (4) & 3040.8 (134) &297.1 (17.6)& 0.102 (23) & 0.146 (2)   & 1.26 (6)    &  3.9 (0.2) \\
 V1928 Aql  & 6486.5032 (8)  &  0.520679 (1) & 4.14 (0.17)& 0.0026 (2) & 4844.4 (144) &  0.5 (5.7) & 0.179 (21) & 0.006 (1)   & 0.49 (3)    & 19.0 (1.5) \\
 V2486 Cyg  & 6497.4738 (20) &  1.272708 (3) & 8.37 (0.60)& 0.0161 (30)& 6558.6 (200) &  4.9 (31.5)& 0.002 (  ) & 0.309 (13)  & 2.34 (30)   &  8.1 (0.9) \\
  \hline
\end{tabular}
\end{table*}

\subsection{HD 24105}

Another target in our sample was HD~24105, which is in fact one component of a visual double named
HEI~9AB (or EPIC~211160717). Its photometric variability with period of about 1.26 days was
discovered by the STEREO satellite \citep{2011MNRAS.416.2477W}, however no detailed analysis of the
binary has been carried out so far.

\begin{figure}
  \centering
  \includegraphics[width=0.48\textwidth]{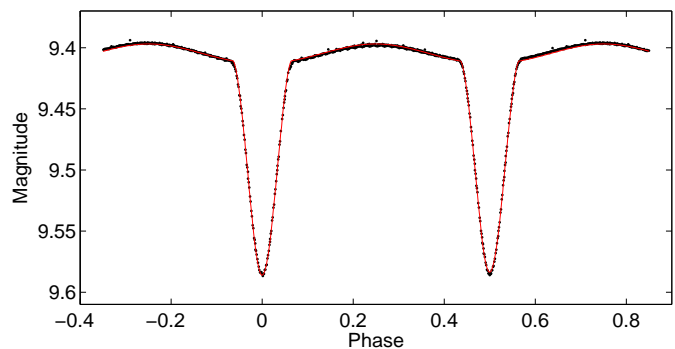}
  \caption{Light curve of HD 24105 based on Kepler data, together with our final fit.}
  \label{FigHD24105LC}
\end{figure}

\begin{figure}
  \centering
  \includegraphics[width=0.48\textwidth]{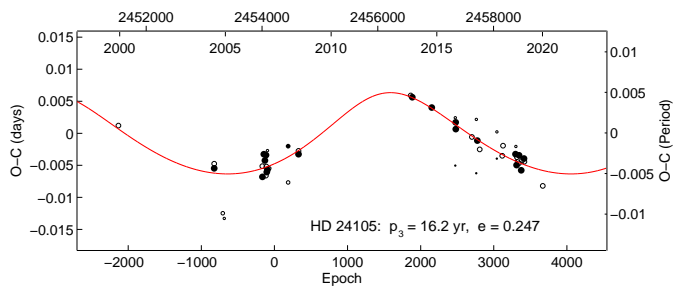}
  \caption{Period analysis of HD 24105.}
  \label{FigHD24105_LITE}
\end{figure}

The two visual components are about 4.4$^{\prime\prime}$ distant, hence the photometry should be
contaminated with the other component. According to the Washington Double Star catalog
(WDS\footnote{http://ad.usno.navy.mil/wds/}; \citealt{2001AJ....122.3466M}), there has been no obvious
movement on the visual orbit of the two components over the last 80 years. According to similar
parallaxes and proper motions from GAIA DR2 \citep{2018A&A...616A...1G}, the pair is really
physical, but with an estimated period of the order of $\sim$kyr.

By far the best quality of the LC is the one by the Kepler satellite. Therefore, we used this
photometry for the LC solution. This fit is presented in Fig.\ref{FigHD24105LC}. The parameters of
the fit are given in Table \ref{TabLC}. As one can see, the fit is of very high quality, both
components are rather detached, moving in circular orbit around each other. Both minima are about
the same depth. The level of the third light resulted in about 20\% of the total light, which set
some constraints on the third-body parameters.

Our analysis of ETV is plotted in Fig.\ref{FigHD24105_LITE}. Parameters of the LITE are also given
in Table \ref{LITEparam}. The apparent variation of its orbital period shows a modulation with
period of about 16~yr, which is slightly more than the time interval covered with data nowadays.
Therefore, we believe our result is reliable, but should be slightly updated when new observations
emerge. The level of the third light is also in accordance with the predicted third-body
hypothesis. Our predicted third body should be detected interferometrically on order-of-magnitude
closer separations than the component already observed at 4.4$^{\prime\prime}$ , and this would be a
challenging task.

\subsection{KU Aur}

The next eclipsing binary in our sample is KU~Aur (= TYC~2422-20-1). Its photometric variability
together with its eclipsing nature and correct orbital period of about 1.3~days were discovered by
\cite{1961AN....286...81P}. Later, the object was observed and its LC analyzed by
\cite{2004IBVS.5499....1L}, revealing its semidetached configuration with a rather deep primary, but
only shallow secondary, eclipse. Its spectral type was most recently given as F8 by the LAMOST
survey \citep{2015RAA....15.1095L}. According to the GAIA catalog, a close
($<1^{\prime\prime}$) visual component also exists.

\begin{figure}
  \centering
  \includegraphics[width=0.48\textwidth]{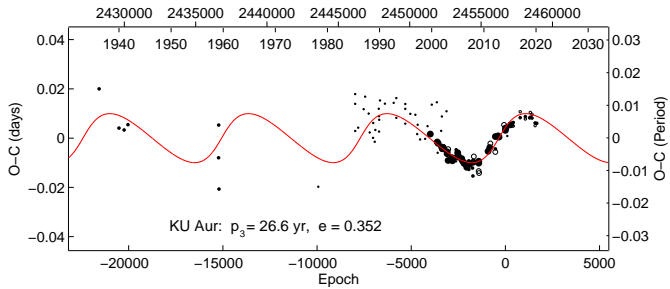}
  \caption{Period analysis of KU Aur.}
  \label{FigKUAur_LITE}
\end{figure}

However, the ETV analysis has not been published so far, despite its extensive collection of eclipse
timings ranging over 80 years. Therefore, we collected all available published minima ($>$ 90
datapoints), and also added the new ones from the SuperWASP, ASAS-SN, and KWS surveys (25 eclipse
timings). With this dataset, we were able to identify the periodic variation in the $O-C$ diagram,
which is plotted in Fig. \ref{FigKUAur_LITE}. The LITE variation is clearly evident, having a period of
about 27~yr and an amplitude of about 14~minutes. The putative third body should have its minimal
mass of about the same as the secondary component, but it should be undetectable interferometrically.
However, the LC analysis published by \cite{2004IBVS.5499....1L} also revealed a small
contribution of the third light, which can be identified with our detected component.

\begin{table*}
 \centering
  \footnotesize
  \caption{The parameters of the LC fits.}  \label{TabLC}
  \begin{tabular}{@{}c c c c c c c c c c c c c c c@{}}
\hline
            &\multicolumn{2}{c}{HD 24105}    &\multicolumn{2}{c}{GH Mon}      &\multicolumn{2}{c}{DK Her}      &\multicolumn{2}{c}{V1134 Her}   &\multicolumn{2}{c}{KIC 6187893} &\multicolumn{2}{c}{V1928 Aql}   &\multicolumn{2}{c}{V2486 Cyg}   \\
   Parameter&Value & Error                   &Value  & Error                  &Value  & Error                  &Value  & Error                  &Value  & Error                  &Value  & Error                  &Value  & Error                  \\
 \hline
  $T_1$ [K] &\multicolumn{2}{c}{5660 (fixed)}&\multicolumn{2}{c}{6300 (fixed)}&\multicolumn{2}{c}{9200 (fixed)}&\multicolumn{2}{c}{6651 (fixed)}&\multicolumn{2}{c}{5854 (fixed)}&\multicolumn{2}{c}{8500 (fixed)}&\multicolumn{2}{c}{8840 (fixed)}\\
  $T_2$ [K] &5655  & 13                      &4163   &  59                    &2503   & 190                    &6320   & 68                     &4699   &  11                    &8505   & 27                     &5679   & 45                     \\  %
  $i$ [deg] &76.50 & 0.08                    &71.99  &  0.32                  &88.04  & 0.26                   &74.77  & 0.59                   &89.47  & 0.16                   &63.79  & 0.32                   &89.62  & 0.34                   \\  %
  $L_1$ [\%]&41.7  & 0.3                     &72.8   &  1.4                   &82.8   & 1.9                    & 37.8  & 1.3                    & 12.6  & 0.2                    & 48.6  & 0.9                    & 39.4  & 0.9                    \\  %
  $L_2$ [\%]&37.6  & 0.4                     &14.6   &  0.9                   &  0.2  & 0.1                    & 30.9  & 1.1                    &  1.6  & 0.1                    & 45.8  & 0.8                    &  5.5  & 0.3                    \\  %
  $L_3$ [\%]&20.7  & 0.7                     &12.6   &  1.2                   &17.0   & 1.2                    & 31.3  & 1.6                    & 85.8  & 0.5                    &  5.6  & 1.2                    & 55.1  & 0.8                    \\  %
  $r_1/a$   &0.229 & 0.002                   &0.277  & 0.004                  &0.195  & 0.005                  &0.453  & 0.003                  &0.338  & 0.001                  &0.384  & 0.007                  &0.339  & 0.002                  \\  %
  $r_2/a$   &0.218 & 0.002                   &0.398  & 0.003                  &0.224  & 0.004                  &0.453  & 0.003                  &0.211  & 0.001                  &0.372  & 0.007                  &0.302  & 0.008                  \\  %
 \hline
\end{tabular}
\end{table*}

\subsection{GU CMa}

Another eclipsing system included in our analysis is GU~CMa (= HD~52721), which has an orbital period
of 1.61 days. This is probably the most interesting system, astrophysically speaking, in our sample.
Due to its nature as an emission line star of spectral type Be (B2Vne, B3e, from
\citealt{2014yCat....1.2023S}), its relatively high brightness of about 6.6~mag in $V$ filter, and its
location near the equator, this star became a quite frequent target for different investigators using
various means. Moreover, it also probably belongs to a star association named CMa~R1. All of these
yielded more than 200 publications on this target, from which surprisingly only very few were on its
eclipsing nature, and none of them were on the ETV analysis. The last and probably the most detailed
analysis of its LC was that one by \cite{2018NewA...59....8S}. They concluded that the system is a
semidetached one, with a secondary evolved from the main sequence.

\begin{figure}
  \centering
  \includegraphics[width=0.48\textwidth]{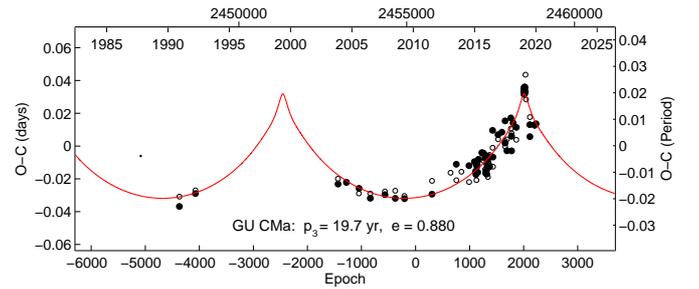}
  \caption{Period analysis of GU CMa.}
  \label{FigGUCMa_LITE}
\end{figure}

We collected all available published times of eclipses from the literature (12 records), to which
we also added nine from our new observations, and, finally, 108 were derived from various photometric
surveys (Hipparcos, ASAS, ASAS-SN, KWS, TESS, and MASCARA). With this huge dataset, we carried out
an analysis, which is plotted in Fig.\ref{FigGUCMa_LITE}. The ETV modulation has a period of
about 19.7 years. As one can see, the last period of the outer orbit is well covered, but the older
observations are missing. Our resulting parameters (see Table \ref{LITEparam}) should hence be
taken as preliminary. However, what can clearly be seen even with our data is a rapid period
change near its periastron passage caused by very high eccentricity. Such a result is noteworthy,
because such a high eccentricity is only known for several systems as it stands.

From the LITE parameters, we can roughly estimate the angular distance of the predicted third
component. This resulted in about 40~mas, which should be detectable with modern techniques.
GU~CMa is also known as a visual double, but its detected companion at a separation of about
0.6$^{\prime\prime}$ is definitely another member of this multiple system.

\subsection{GH Mon}

The system GH~Mon (= GSC~04818-02919) was discovered as a variable by \cite{1968AAHam...7..381W}. Its
orbital period is of about 0.7~days and it is one of the only seldom-investigated stars. No detailed
analysis of this system has been published so far: only several eclipse times. Its spectral type was
classified as F6 from LAMOST by \cite{2015RAA....15.1095L}. Again, from the GAIA catalog, it has a
close ($\approx 2^{\prime\prime}$) visual component.

\begin{figure}
  \centering
  \includegraphics[width=0.48\textwidth]{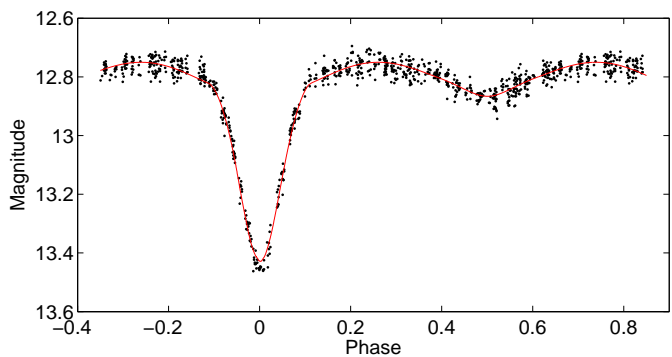}
  \caption{Light curve of GH Mon based on ASAS-SN data.}
  \label{FigGHMonLC}
\end{figure}

Therefore, we performed an LC analysis using the {\sc PHOEBE} code and the ASAS-SN data,
which seemed to be of the best quality among the other photometric surveys for this star. This
solution is presented in Fig. \ref{FigGHMonLC}, while its parameters are given in Table
\ref{TabLC}. The orbit of eclipsing components is circular, with the secondary component being slightly
evolved from the main sequence, but still in a detached configuration.

\begin{figure}
  \centering
  \includegraphics[width=0.48\textwidth]{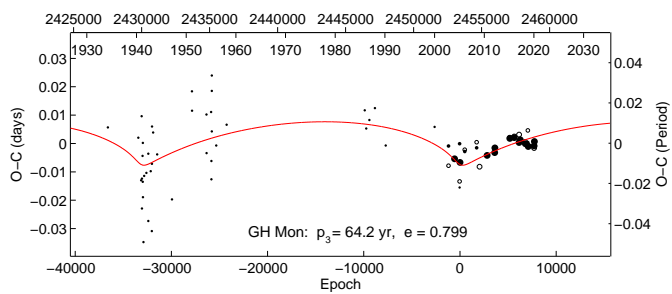}
  \caption{Period analysis of GH Mon.}
  \label{FigGHMon_LITE}
\end{figure}

For the ETV analysis, we used all archival data together with our new derived ones. The final $O-C$
diagram with its periodicity of about 64 yr is plotted in Fig.\ref{FigGHMon_LITE}, and the
parameters of the fit are given in Table \ref{LITEparam}. Obviously, the outer period is too long,
and the coverage is still rather poor. However, the last minima times since 2005 show clear
evidence of the period change. What is interesting is the fact that even after subtraction of this
LITE fit, there also remains a small additional
variation with smaller amplitude
and shorter pseudo-period on the residuals. However, any statement about a possible fourth body in the system would still
be rather premature.

\subsection{AZ Vel}

The star named AZ~Vel has also been rather neglected in binary star research. The discovery was made
by \cite{1939AN....268..165G}, who also gave its proper orbital period of about 0.776 days and an
Algol-type LC classification. However, since then no detailed analysis had been carried out until our
recent paper on AZ~Vel, which was based on the photometric data from the INTEGRAL/OMC
\citep{2010NewA...15..150Z}. The result was that the system is a detached one, but its spectral
type and primary temperature is still rather uncertain (the system's spectrum has not yet been obtained in
any publication).

\begin{figure}
  \centering
  \includegraphics[width=0.48\textwidth]{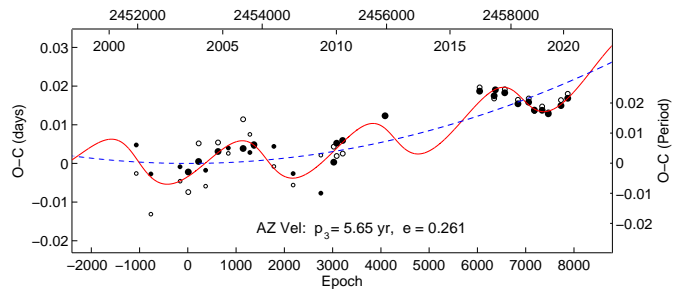}
  \caption{Period analysis of AZ Vel, showing slow period increase.}
  \label{FigAZVel_LITE}
\end{figure}

We collected all available times of eclipses and enlarged the dataset with the new ones from the
OMC, ASAS, and ASAS-SN surveys. On this collection of more than 60 minima covering 20 years, we
performed an ETV analysis. The results can be seen in Fig. \ref{FigAZVel_LITE} and clearly show the
ETV variation with a periodicity of 5.65 yr. AZ~Vel seems to be the only system where
quadratic ephemerides were also used besides the LITE hypothesis. Its interpretation is still an open
question due to the fact that we are dealing with a detached configuration of the eclipsing double.

 \subsection{DI Lyn}

The system named DI~Lyn (or also A~Hya = HD~82780) is an Algol-type eclipsing binary discovered
originally on the Hipparcos data \citep{1997A&A...323L..49P}. Later, \cite{1998IBVS.4649....1W}
analyzed the $BVRI$ LCs together with the radial velocities resulting in a set of physical
parameters of both components of F2V spectral type. Moreover, additional more
distant components also exist in the system, making the whole system at least a quintuple one. Besides the
eclipsing 1.7-day component A, component B also exists, with a 28-day period detected
spectroscopically. Quite recently, \cite{2019AJ....158..167T} published the solution of the visual
orbit of the Aa-Ab subsystem, resulting in an orbital period of 50.34~yr. The last additional component at
about a 25$^{\prime\prime}$ distance is also a physical one according to its parallax and proper
motion. We invite the reader to consult the paper by \cite{2006A&A...450..681T} summarizing the whole architecture of the
system.

\begin{figure}
  \centering
  \includegraphics[width=0.48\textwidth]{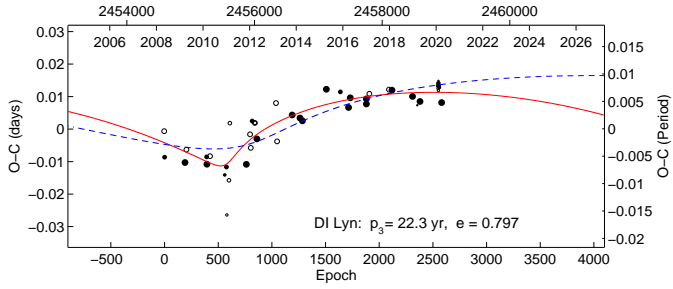}
  \caption{Period analysis of DI Lyn, compared with orbit from \cite{2019AJ....158..167T}.}
  \label{FigDILyn_LITE}
\end{figure}

Due to the fact that the LC had already been analyzed, we only collected all available times of eclipses
for the proper ETV analysis. Apart from our observations of eclipses, we also used the TESS data to
derive new eclipse times. Therefore, we analyzed the data spanning over 12 years. With these
data, we carried out an ETV analysis on DI~Lyn. Our analysis showed that the period of an ETV
modulation is of about 22 yr, and the rest of the parameters are given in Table \ref{LITEparam},
while the plot with the LITE fit is given in Fig. \ref{FigDILyn_LITE}.

Due to the fact that the orbit of the Aa-Ab pair had already been analyzed, we also tried to identify the ETV
detected in our data with this 50.34-yr orbit. Such a comparison is also plotted in
Fig.\ref{FigDILyn_LITE} as a blue dashed curve, where we fixed the orbital parameters given by
\cite{2019AJ....158..167T} and computed only the ephemerides of the 1.7-day eclipsing binary and
the LITE amplitude. The periastron passage of both these solutions seem to be close to each other,
indicating that this ETV variation really belongs to the Aa-Ab orbit. However, fixing the orbital
parameters from \cite{2019AJ....158..167T} to our dataset, we get about a 46\% worse sum of square
residuals than with all of our parameters that were released freely and converged into the solution
presented in Table \ref{LITEparam}. Such a discrepancy should be studied in upcoming years:
observing the target year-by-year and following the ETV curve.

 \subsection{DK Her}

The eclipsing system DK~Her (or HD~155700), with an orbital period of about 1.9 days, shows very
different depths of both eclipses. Despite the fact that its eclipsing nature has been known for many
decades, an analysis of the star is still missing. Neither LC analysis nor period analysis can be found
in published papers concerning the star. The only piece of information is a spectrum from LAMOST
\citep{2015RAA....15.1095L}, where the authors gave the type of A1IV. DK~Her was also found to be a
visual binary. The two components separated by about 10$^{\prime\prime}$ on the sky, are physical
according to GAIA (similar parallax and proper motion).

\begin{figure}
  \centering
  \includegraphics[width=0.48\textwidth]{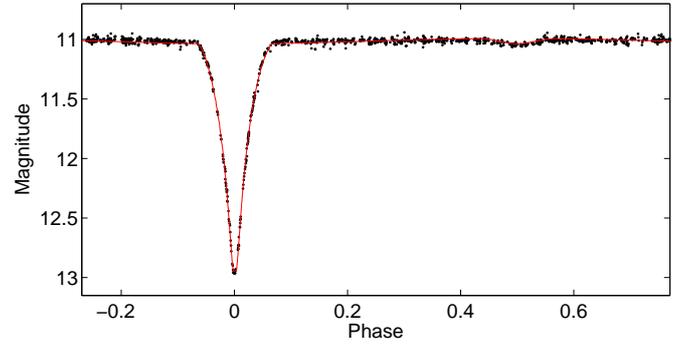}
  \caption{Light curve of DK Her based on ASAS-SN data.}
  \label{FigDKHerLC}
\end{figure}

Therefore, we performed an LC as well as an eclipse timing analysis. To study the LC, we used
the data from the ASAS-SN survey, and particularly the $g$-band filter. The result of our analysis is
plotted in Fig. \ref{FigDKHerLC}. As one can see, the luminosity ratio is very extreme here, which is
also seen at the shape of the LC, with an almost invisible secondary minimum.

\begin{figure}
  \centering
  \includegraphics[width=0.48\textwidth]{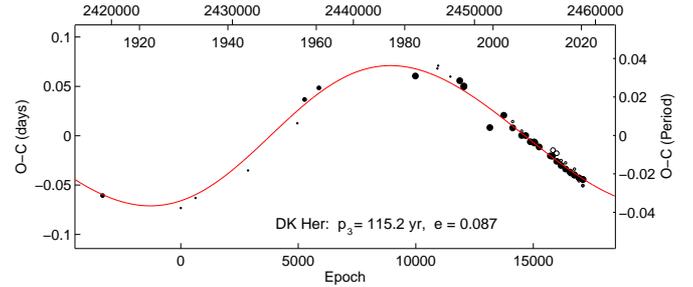}
  \caption{Period analysis of DK Her.}
  \label{FigDKHer_LITE}
\end{figure}

The analysis of eclipse times collected over the last century is plotted in Fig. \ref{FigDKHer_LITE}.
Due to very deep primary eclipses, we also believe that the old visual and photographic minima times
are reliable enough to be used in the ETV analysis. The parameters of our fit are given in Table
\ref{LITEparam}, showing that this is the system with the longest orbital period in our sample of stars
(115~yr). However, despite the period of a putative third body being long, it is almost covered with
data nowadays.

 \subsection{GQ Dra}

The system named GQ~Dra (= HD~158260 = HIP~85277) is a variable of the $\beta$~Lyrae type with an
orbital period of 0.766 days, and it was discovered from the Hipparcos data. The first study on this star was
the one of \cite{2000IBVS.4988....1A}, but this gave no detailed analysis of its LC. A very detailed
analysis was published later by \cite{2015AJ....150..193Q}, revealing that the star is a classical
semidetached system with secondary filling its respective Roche lobe. The authors speculated about
a possible mass transfer between the components due to their period analysis of eclipse times. Its
spectral type was derived as A3 \citep{1975ascp.book.....H}, which is quite typical for this type of
semidetached stars. The LAMOST survey published its type as A6IV \citep{2015RAA....15.1095L}.
Moreover, the system GQ~Dra is also known as a visual binary, with both components
4$^{\prime\prime}$ away from each other, and no mutual orbital motion can be seen during the last century
\citep{2001AJ....122.3466M}. Also, no astrometric acceleration can be seen in the proper motion data
\citep{2018ApJS..239...31B}. However, such an effect caused by our LITE orbit depends on the masses
and orientation of the orbit toward the observer.

\begin{figure}
  \centering
  \includegraphics[width=0.48\textwidth]{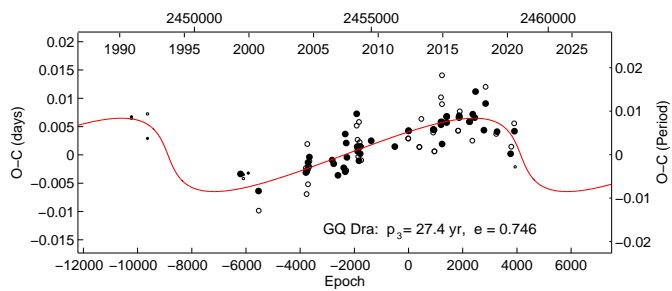}
  \caption{Period analysis of GQ Dra.}
  \label{FigGQDra_LITE}
\end{figure}

Because of the fact that the LC was analyzed in detail by \cite{2015AJ....150..193Q}, we decided only to
perform the ETV analysis of the minima timings. Most of these data were taken from the already
published minima compilations. We also added new eclipse times derived from the
Hipparcos, NSVS, and SWASP photometry. Besides that, several were added from our
dedicated ground-based observations. This dataset covers a time span of 30 years, and the result
of our analysis is plotted in Fig. \ref{FigGQDra_LITE}. The parameters are given in Table
\ref{LITEparam}, with the amplitude of LITE only being about 9.5 minutes, and the period about
27.4~yr. The rapid period change close to the year 2020 is evident now, indicating rather high
eccentricity of the orbit. All of these findings should be confirmed later in the upcoming period
of the outer orbit.

 \subsection{V624 Her}

The 3.89-d eclipsing system V624~Her (= HD~161321 = HIP~86809) is the brightest star in our sample.
Due to its magnitude, it has been studied quite often in the past. Besides that, it was also classified
as a metallic star \citep{1984AJ.....89.1057P}, with the primary component even having the largest
radius among the known Am stars with known properties. \cite{1984AJ.....89.1057P} derived the basic
physical parameters of the system, as well as both components based on adequately good photometry
and spectroscopy. Moreover, the star is also known as a visual double, but with both components
rather distant on the sky ($\approx$40$^{\prime\prime}$). According to the GAIA values it is probably
optical.

\begin{figure}
  \centering
  \includegraphics[width=0.48\textwidth]{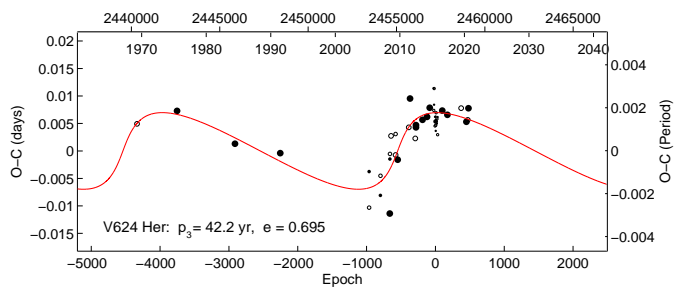}
  \caption{Period analysis of V624 Her.}
  \label{FigV624Her_LITE}
\end{figure}

Owing to the already published solution of LC and RV curves by \cite{1984AJ.....89.1057P}, we
decided only to carry out a detailed ETV analysis of V624~Her. We collected all available times
of eclipses from the literature, and we added several more from the ASAS and MASCARA surveys, and
the satellite Hipparcos. This dataset covers more than 50 years of eclipse time monitoring. The result
of our fitting is plotted in Fig. \ref{FigV624Her_LITE}, with a periodicity of about 42 years,
while the parameters are given in Table \ref{LITEparam}. The distant visual component is definitely
different to the one detected with LITE. Such a closer component (with predicted angular separation
of about 140~mas; see Table \ref{LITEparam}) could possibly be detected via interferometry due to
the high level of brightness of the star (eclipsing the binary itself), however the problematic issue would be the
luminosity ratio (low brightness of the third component).

 \subsection{V1134 Her}

\begin{figure}
  \centering
  \includegraphics[width=0.48\textwidth]{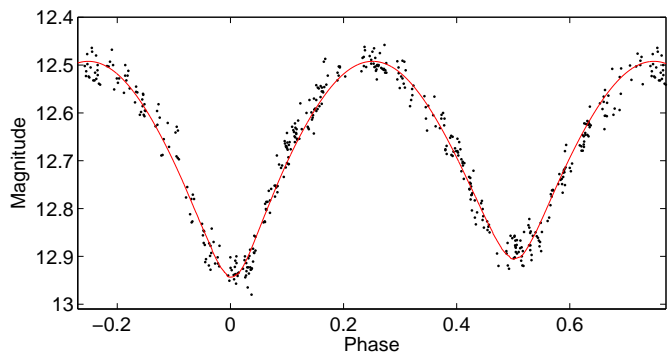}
  \caption{Light curve of V1134 Her based on ASAS-SN data.}
  \label{FigV1134HerLC}
\end{figure}

The star V1134~Her (= GSC~01031-01766) is probably the least studied system in our compilation. It
is a contact-eclipsing binary with an orbital period of about 0.6~days, however no detailed
analysis about the star has been published so far. Concerning its spectral type, it is rather uncertain. No
spectral observation exists, hence we can only estimate its type from the photometric indices.
Some $(B-V)$ indices from APAS \cite{2015AAS...22533616H}, or UCAC4 \cite{2013AJ....145...44Z}
range from 0.59 - 0.61~mag, yielding spectral types of about G0-1V. On the other hand, the
$(J-H)$ values from 2MASS \citep{2006AJ....131.1163S} would better fit spectral types of about
F7V. Therefore, we decided to use the temperature estimation based on the GAIA DR2 by
\cite{2019AJ....158...93B}, which gives $T_{eff} = 6651$~K for the LC analysis. The photometric
analysis was carried out on the data provided by ASAS-SN \citep{2014ApJ...788...48S} and its $V$
filter. For the results of the LC fitting, see Table \ref{TabLC} and the final fit presented in
Fig. \ref{FigV1134HerLC}. As one can see, the third component significantly contributes to the
total luminosity of the whole system.

\begin{figure}
  \centering
  \includegraphics[width=0.48\textwidth]{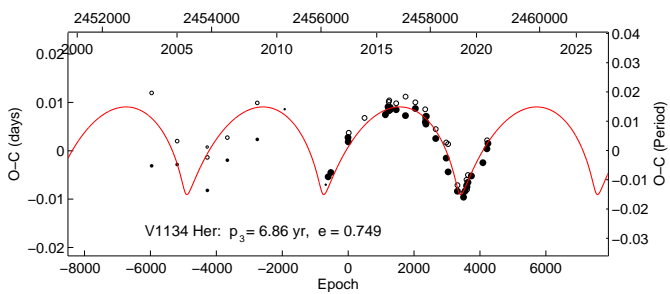}
  \caption{Period analysis of V1134 Her.}
  \label{FigV1134Her_LITE}
\end{figure}

The ETV analysis of eclipse times is presented in Fig. \ref{FigV1134Her_LITE}. Most of these data
were derived from the surveys ASAS and ASAS-SN, and several new ones also from our dedicated
observations. The period variation caused by LITE is rather fast, having a period of only about
6.9~years. Its coverage with current data is sufficient for such an analysis, but new observations
in the coming years would be of great benefit to confirm our hypothesis with higher
conclusiveness.

 \subsection{KIC 6187893}

Only one system from our sample was discovered with Kepler satellite: KIC~6187893 (=
TYC~3128-1653-1). The star shows a rather detached LC with an orbital period of about 0.8~
days. Its spectral type was derived as F7 from the LAMOST survey data \citep{2015RAA....15.1095L},
but its effective temperature was estimated as 5854~K \citep{2019ApJS..244...43Z}.

\begin{figure}
  \centering
  \includegraphics[width=0.48\textwidth]{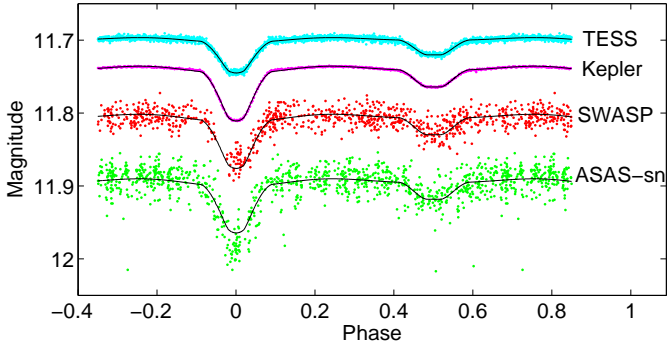}
  \caption{Light curves of KIC 6187893 showing the comparison of quality of different data sources.}
  \label{FigKIC6187893LC}
\end{figure}

Therefore, we fixed this temperature as the effective temperature of the primary component and
carried out the LC analysis. We decided to show (Fig.\ref{FigKIC6187893LC}) our fit on the
best-quality data from Kepler compared with the other data from different photometric
surveys like SWASP, ASAS-sn, or TESS. Not only are there different scatters, but also different eclipse
depth is clearly seen in these plots. This effect is caused by the varying angular resolution
(pixel size) of the particular survey telescope. The results of our LC fitting are given in Table
\ref{TabLC} (results given for the most precise LC fit on the Kepler data).

\begin{figure}
  \centering
  \includegraphics[width=0.48\textwidth]{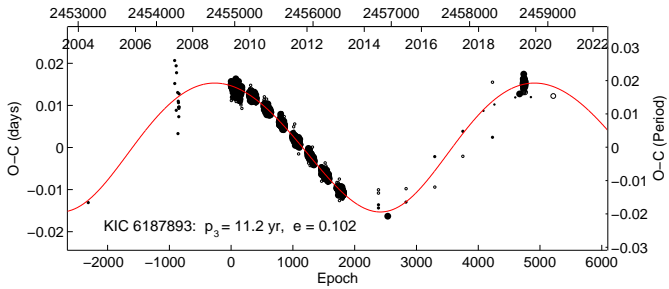}
  \caption{Period analysis of KIC 6187893.}
  \label{FigKIC6187893_LITE}
\end{figure}

The analysis of minima times is plotted in Fig. \ref{FigKIC6187893_LITE}. The periodic signal is
clearly visible there, with a periodicity of about 11~yr. The data from the Kepler satellite
were the richest among those available, as well as being of the highest quality. However, the other
photometric observations nicely completed the whole third-body period, which is now fully covered
with data. From the very large fraction of the third light from the LC solution, it appears that
the third-body orbit must be very inclined to produce such a large light contribution with such a low
LITE amplitude.

 \subsection{V1928 Aql}

The next system under our analysis is V1928~Aql (= HD~180848 = TYC~1042-1657-1), which had never been
studied before as an eclipsing binary, but which has already been known as a visual double star for a
long time. Its two components are about 7$^{\prime\prime}$ distant from each other. Its
photometric variability with a period of about 0.52 days was discovered by our team
\citep{2017IBVS.6204....1Z}, and since then has been quite frequently observed.
\cite{1957ApJ...125..195N} included the star on their list of carbon stars in the Milky Way, however no
other more recent information about this concern has since been published. The classification of the
system is currently given as A3 \citep{1975ascp.book.....H}.

\begin{figure}
  \centering
  \includegraphics[width=0.48\textwidth]{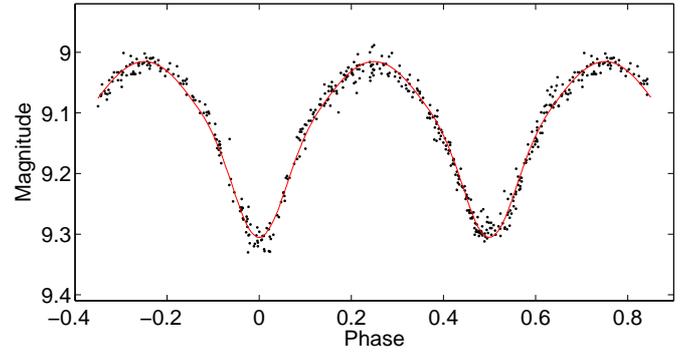}
  \caption{Light curves of V1928 Aql based on ASAS data.}
  \label{FigV1928AqlLC}
\end{figure}

\begin{figure}
  \centering
  \includegraphics[width=0.48\textwidth]{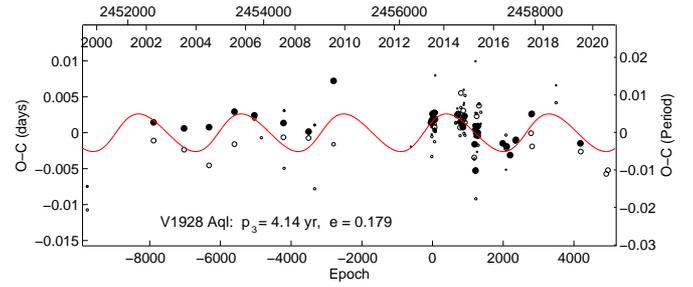}
  \caption{Period analysis of V1928 Aql.}
  \label{FigV1928Aql_LITE}
\end{figure}

For the LC analysis, we used the ASAS photometry in $V$ filter, which seems to be of the best
quality due to the relatively high brightness of the star. The result of our fitting is plotted in Fig.
\ref{FigV1928AqlLC}. The system is a contact one with very similar components, and the third
light value is very small (see Table \ref{TabLC} with the parameters). Such a finding should
be confronted with the results from our ETV analysis. These results are given in Table
\ref{LITEparam}, while the plot is given in Fig. \ref{FigV1928Aql_LITE}. V1928~Aql shows the lowest
amplitude of LITE among the stars in our sample, and it has an ETV period of about 4.14 years. This is
the reason why the plot given in Fig. \ref{FigV1928Aql_LITE} seems to be rather scattered.
However, the scatter of the older ASAS data from between 2002 and 2010 is caused by our method of minima
derivation, by which the whole season of observations is used to construct the seasonal LC subsequently analyzed. Even during one season, there is an apparent period change blurring the
shape of the LC.

 \subsection{V2486 Cyg}

The last system in our list of analyzed binaries is V2486~Cyg (= GSC~03173-01826 =
BD+39~4506). It was discovered as a variable by \cite{1962IBVS...17....1S}, but since then no
analysis of the system has been published except for the times of eclipses. It is an Algol-type binary
with an orbital period of about 1.27~days. It is also known as a visual double pair, but with no
obvious mutual motion of the two components, which are about 20$^{\prime\prime}$ from each other.
However, due to quite different parallaxes and proper motions of both components
\citep{2018A&A...616A...1G}, their physical relationship is rather questionable.

\begin{figure}
  \centering
  \includegraphics[width=0.48\textwidth]{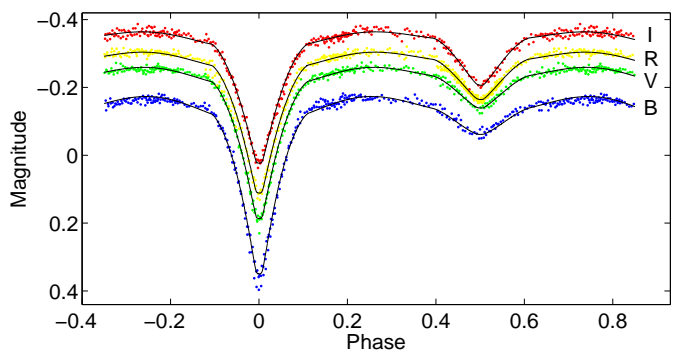}
  \caption{Differential $BVRI$ light curves of V2486 Cyg.}
  \label{FigV2486CygLC}
\end{figure}

Due to missing LC analysis, we decided to include our solution in this study. Its spectral type
had been classified as either A2 \citep{1975ascp.book.....H} or A0 \citep{1964MitVS...2...85G}. Therefore, we
fixed its primary temperature as 8840~K. The final fit of our own data obtained in standard
Johnson-Cousins $BVRI$ filters is plotted in Fig. \ref{FigV2486CygLC}, while the parameters are
given in Table \ref{TabLC}. In the table, the fractional luminosities are given for the $V$ filter.
These values significantly differ between different filters: in the blue region, the primary
component is almost as luminous as the tertiary (45\%:52\%), while in the red band the third
component significantly dominates (35\%:56\%). From these numbers, it can clearly be seen that
the putative third component is much more red than the primary, and it probably also dominates the
luminosity of the whole system.

\begin{figure}
  \centering
  \includegraphics[width=0.48\textwidth]{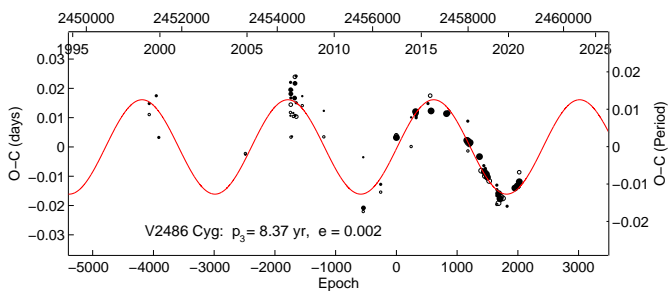}
  \caption{Period analysis of V2486 Cyg.}
  \label{FigV2486Cyg_LITE}
\end{figure}

Such a finding is well supported by our ETV analysis. From the available eclipse times, it
can clearly be seen (Fig. \ref{FigV2486Cyg_LITE}) that the period variation with a periodicity of
about 8.4~yr is undoubtable. From this solution, we derive a minimal third-body mass of at least
2.3~$\mathrm{M}_\odot$, which is more than that of the primary component. Such a result is in good
agreement with the high third light value from the LC solution.

Moreover, during our photometric monitoring of the system V2486~Cyg, we discovered that the close-by
target \object{TYC 3173-01309-1} is also a variable. Its type is W~UMa, and its period is of about
0.3952~days.

\section{Conclusion}

We carried out a thorough analysis of period variations for 14 new eclipsing systems showing
apparent period changes. These changes can be attributed to hidden distant components in these
systems, making them even more interesting multiples. For cases like DI~Lyn (which was up to now
classified as a quintuple) such a finding is of high importance when studying its long-term
stability.

The powerful method of ETV analysis still brings some benefits to modern stellar astrophysics. It is so simple that even non-professional astronomers (like the co-authors of the present
study) can significantly contribute with their small or moderate-sized telescopes. The
method can also detect the third-body periods that are otherwise barely detectable. The mutliple-system
statistics of outer periods still have two observational peaks: in shorter periods, due to
spectroscopic triples, and the longer-period ones are detected via interferometry or direct imaging
(see e.g., \citealt{2008MNRAS.389..925T}). The ETV analysis of eclipsing binaries can serve as
an effective method for bridging the gap between these two peaks. However, for a final confirmation of
these suspected triples, one needs precise spectroscopy in order to detect them directly in the spectra, or
at least to reveal a change in the systemic velocity of the binary itself.

\begin{acknowledgements}
This work has made use of data from the European Space Agency (ESA) mission {\it Gaia}
(\url{https://www.cosmos.esa.int/gaia}), processed by the {\it Gaia} Data Processing and Analysis
Consortium (DPAC, \url{https://www.cosmos.esa.int/web/gaia/dpac/consortium}). Funding for the DPAC
has been provided by national institutions, in particular the institutions participating in the
{\it Gaia} Multilateral Agreement.
     We would also like to thank the Pierre Auger Collaboration for
the use of its facilities. The operation of the robotic telescope FRAM is supported by the grant of
the Ministry of Education of the Czech Republic LM2018102. The data calibration and analysis
related to the FRAM telescope is supported by the Ministry of Education of the Czech Republic
MSMT-CR LTT18004 and MSMT/EU funds CZ.02.1.01/0.0/0.0/16$\_$013/0001402.
   We also thank the ASAS, SWASP, CRTS, TESS, MASCARA, ZTF, and Kepler teams for making all of the
observations easily available.
    This paper makes use of data from the DR1 of the WASP data \citep{2010A&A...520L..10B}  provided by
the WASP consortium, and the computing and storage facilities at the CERIT Scientific Cloud, reg.
no. CZ.1.05/3.2.00/08.0144, which is operated by Masaryk University, Czech Republic.
    The CSS survey is funded by the National Aeronautics and Space Administration under Grant No.
NNG05GF22G issued through the Science Mission Directorate Near-Earth Objects Observations Program.
The CRTS survey is supported by the U.S.~National Science Foundation under grants AST-0909182 and
AST-1313422.
    Work is based on the data from the OMC Archive at CAB (INTA-CSIC), pre-processed by ISDC.
   We would also like to thank Mr. Peter Nos\'a\v{l} for using his photometric data of V1134~Her.
This research has made use of the SIMBAD and VIZIER databases, operated at CDS, Strasbourg, France
and of NASA Astrophysics Data System Bibliographic Services.
\end{acknowledgements}

\end{document}